\documentclass{article}

\def\nn{\nonumber}
\def\ket#1{| #1 \ra }
\def\bra#1{\la #1}

\def\la{\langle}
\def\ra{\rangle}

\def\l{\left}
\def\r{\right}
\def\nn{\nonumber}

\def\beq{\begin{equation}}
\def\eeq{\end{equation}}
\def\bea{\begin{eqnarray}}
\def\eea{\end{eqnarray}}
\def\barr{\begin{array}}
\def\earr{\end{array}}

\def\ln#1{\log{\l( #1 \r)}}

\begin{document}

\begin{titlepage}
\begin{flushright}
Roma-1357/03\\
SHEP 0323\\
SISSA 69/03 EP\\
\end{flushright}
\vskip 0.5cm
\begin{center}
{\Large \bf Finite-Volume  Partially-Quenched  Two-Pion Amplitudes \\
\vskip0.2cm in the \boldmath{$I=0$} Channel} \vskip1cm {\large\bf
C.-J.D.~Lin$^a$, G.~Martinelli$^b$, E.~Pallante$^c$,
C.T.~Sachrajda$^{a}$, G.~Villadoro$^b$}\\ \vspace{.5cm}
{\normalsize {\sl $^a$ School of Physics and Astronomy, Univ. of
Southampton,\\ Southampton, SO17 1BJ, UK. \\ \vspace{.2cm} $^b$
Dip. di Fisica, Univ. di Roma ``La Sapienza" and INFN,\\ Sezione
di Roma, P.le A. Moro 2, I-00185 Rome, Italy.}}\\ \vspace{.2cm}
$^c$ {\sl SISSA and INFN, Sezione di Trieste, Via Beirut 2-4,
34013, Trieste, Italy.}

\vskip1.0cm {\large\bf Abstract:\\[10pt]} \parbox[t]{\textwidth}{{
We present a study of the finite-volume two-pion matrix elements
and correlation functions of the $I=0$ scalar operator, in full
and partially quenched QCD, at one-loop order in chiral
perturbation theory. In partially quenched QCD, when the sea and
valence light quark masses are not equal, the lack of unitarity
leads to the same inconsistencies as in quenched QCD and the
matrix elements cannot be determined. It is possible, however, to
overcome this problem by requiring the masses of the valence and
sea quarks to be equal for the $u$ and $d$ quarks while keeping
the strange quark ($s$) quenched (or partially quenched), but only
in the kinematic region where the two-pion energy is below the
two-kaon threshold. Although our results are obtained at NLO in
chiral perturbation theory, they are more general and are also
valid for non-leptonic kaon decays (we also study the matrix
elements of $(8,1)$ operators, such as the QCD penguin operator
$Q_6$). We point out that even in full QCD, where any problems
caused by the lack of unitarity are clearly absent, there are
practical difficulties in general, caused by the fact that
finite-volume energy eigenstates are linear combination of
two-pion, two-kaon and two-$\eta$ states. Our work implies that
extracting $\Delta I=1/2,\ K\to\pi\pi$ decay amplitudes from
simulations with $m_s=m_{d,u}$ is not possible in partially
quenched QCD (and is very difficult in full QCD).}}
\end{center}
\end{titlepage}

\section{Introduction}\label{sec:intro}
Several methods have been proposed to compute non-leptonic kaon
decay amplitudes in lattice QCD.  In principle one could determine
physical amplitudes, including the final state interaction  (FSI)
phase shifts, from matrix elements of the relevant
operators of the effective Hamiltonian, computed on the lattice in
full QCD, with realistic quark masses, in a finite
volume~\cite{LL,noi}. This programme requires, however, computer
resources which are not available at present, and which will not
be available in the foreseeable future. For this reason several
strategies for the determination of the matrix elements at
next-to-leading order (NLO) in the chiral expansion have been
presented in the literature~\cite{spqr}-\cite{Laiho:2003uy}. The
general idea is to compute matrix elements at unphysical quark
masses and for a range of energies and momenta for the mesons in
order to determine all the low-energy constants required at NLO in
Chiral Perturbation Theory ($\chi$PT). $\chi$PT at NLO can then be
used to obtain the physical matrix elements.

A number of convenient choices have been suggested for the
unphysical kinematics at which the matrix elements can be computed
directly, and from which the complete set of couplings (low-energy
constants) of the weak chiral Lagrangian at NLO in $\chi$PT can
subsequently be determined. The original proposal was to compute
the $K\to\pi\pi$ matrix elements with the kaon and one of the
final state pions at rest, and to vary the momentum of the second
pion (SPQR kinematics)~\cite{spqr}. Laiho and Soni (LS) on the
other hand, propose to combine the $K \to 0$, $K \to \pi$ and $K
\to \pi \pi$ amplitudes, with the two pions at
rest~\cite{Laiho:2002jq,Laiho:2003uy}. In this case one has to
isolate only the contribution from the two-pion state with the
smallest energy in the correlation functions. LS also argue that
this method can be used in partially quenched two-flavour QCD
(PQ2), provided that the mass of the ``sea'' quarks in the loops,
$m_{S}$, is the same as the ``valence'' light quark mass,
$m_{V}=m_{u,d}$ (the strange quark is
quenched)~\cite{Laiho:2003uy}.

In this paper we present a study of two-pion amplitudes in the
$I=0$ channel and the corresponding correlation functions in a
finite volume in partially quenched QCD. The results discussed
below are important for recent attempts to develop new strategies
for computing $K\to\pi\pi$ decay amplitudes using numerical
simulations on the lattice. We also comment on subtleties in the
evaluation of two-pion matrix elements in full QCD.

We have previously used $\chi$PT at NLO to study the determination
of matrix elements with two-pion external states in quenched QCD
both for $I=2$ final states (for which the difficulties discussed
below are not present)~\cite{Lin:2002nq} and for $I=0$ final
states~\cite{qi0}. For $I=0$ final states, we had found that the
absence of unitarity in quenched QCD means that the calculation of
the amplitudes is not possible~\cite{qi0}. The LS proposal has
prompted us to extend these studies to partially quenched
theories. In this paper we confirm that the inconsistencies of the
quenched theory, and in particular those resulting from the mixing
of hadronic and ghost states, are also generally present in
partially quenched QCD, unless the masses of the sea and valence
quarks are degenerate. In PQ2, the masses of the sea and valence
$u$ and $d$ quarks are indeed degenerate, the $\eta^{\prime}$ is
heavy and decouples from the low energy Lagrangian, nevertheless
we show that the lack of unitarity prevents the extraction of
``physical'' amplitudes from finite volume correlation functions
when the two-pion energy $W_\pi\ge W^{min}_{K}\simeq 2 M_{K}$,
where $W^{min}_{K}$ is the smallest two-kaon energy in the finite
volume, corresponding to the two-kaon threshold (throughout this
paper we assume that $W_\pi$ is sufficiently small that
inelasticity effects due to states with more than two mesons can
be neglected). We illustrate this point with explicit calculations
in $\chi$PT of the correlation function of the scalar operator $S$
with two pion fields (we had used the same example in our study of
quenched QCD~\cite{qi0}). In addition we also consider $(8,1)$
operators, such as the QCD penguin operator $Q_{6}$ whose matrix
elements are important in the theoretical prediction for
$\varepsilon^\prime/\varepsilon$, and confirm the same features.

Our work clarifies the conditions necessary to evaluate matrix
elements with two-pion external states, at NLO in $\chi$PT, in
partially quenched QCD. In particular, in PQ2 the necessary
condition $W_\pi< W^{min}_{K}$ means that it is not possible to
extract the $K \to \pi \pi$ matrix elements when
$m_{S}=m_{V}=m_{s}$, where $m_{s}$ is the mass of the strange
quark. Since these matrix elements are needed for the
implementation of the LS proposal~\cite{Laiho:2003uy}, we conclude
that it cannot be applied in PQ2.

Our work also shows that, in spite of the difficulties mentioned
above, the two-flavour partially quenched theory, PQ2, can still
be useful at least at NLO in the chiral expansion. We show that a
consistent extraction of the ``physical'' matrix elements is
obtained if we work with $m_{K} > m_{\pi}$ ($m_{s} > m_{V} =
m_{S}$), at an energy of the two-pion state in the finite volume,
$W_{\pi}$, which satisfies the Lellouch-L\"uscher (LL) elasticity
condition $W_{\pi} < W^{min}_{K}\simeq 2 M_{K}$. The consistency
of PQ2 with this choice of the parameters opens new avenues for
the calculation of the $\Delta I=1/2$ rule and
$\varepsilon^{\prime}/\varepsilon$ amplitudes from $K \to \pi \pi$
matrix elements following the proposal of
refs.~\cite{spqr,Lin:2002nq}. For these transitions it would be
very difficult in practice to determine the amplitudes from a full
QCD three-flavour computation of the $K\to\pi\pi$ matrix elements,
with the mass of the strange quark different from the light quark
masses. A two flavour unquenched calculation, although difficult,
is nevertheless, feasible in the near future. With the required
kinematical conditions however, one loses the practical advantage
of using only pions at rest~\footnote{Alternatively, one can work
with $m_{K} \neq W_{\pi}$ and additional ultraviolet power
divergences have to be subtracted in the matrix element.}. Below,
we will also comment on some of the subtleties in the extraction
of the $K \to \pi \pi$ amplitude with $m_{K} = m_{\pi}$ in full
QCD.

We make the further important, if somewhat disappointing,
observation. In ref.~\cite{sh2} it had been suggested that it is
possible to obtain {\it exact} physical information about full QCD
from three-flavour partially quenched QCD (PQ3), i.e.~from
unquenched simulations with three flavours at $m_{S} \neq m_{V}$.
The practical advantage of PQ3 is that it is computationally less
expensive to vary, and in particular to reduce, the masses of the
valence quarks. Having determined the low-energy constants of
$\chi$PT in PQ3, one can then extrapolate to the physical point,
$m_{S} = m_{V}$. We have explicitly checked that for the cases at
hand with two-pion final states (the matrix element of the scalar
operator or kaon decay amplitudes) this suggestion does not work
and one has the same difficulties as in PQ2: in infinite volume
the amplitude is singular at threshold, there is mixing of
hadronic and ghost states and the finite-volume correlation
functions are plagued by terms which grow linearly or cubically
with the volume thus making the extraction of physical amplitudes
impossible. Moreover, even at fixed finite volume, the presence of
terms which depend quadratically or cubically on the time
distances prevents the reliable determination of the matrix
element. Perhaps for some simple quantities, such as the leptonic
decay constants of mesons or semileptonic form factors, the
strategy proposed in~\cite{sh2} can be used. From the above
considerations, we conclude that this is not true in general.

The plan for the remainder of this paper is as follows. In the
following section we discuss the extraction of the matrix elements
in full QCD. We show that although the theory is unitary, for the
degenerate case $m_s=m_d=m_u$ the fact that in a finite volume the
energy eigenstates are linear combinations of two-pion, two-kaon
and two-$\eta$ states makes the extraction of two-pion matrix
elements subtle and difficult. We then turn to partially quenched
QCD for which the explicit results at NLO in $\chi$PT for both the
scalar operator and for $(8,1)$ operators are presented in the
appendix. In section~\ref{sec:partially} we study the implications
of these results for the determination of two-pion matrix elements
in partially quenched QCD. Section~\ref{sec:concs} contains our
conclusions.

\section{Two-pion matrix elements in full QCD  with 
\boldmath{$m_{K} > m_{\pi}$}
and \boldmath{$m_{K} = m_{\pi}$}} \label{sec:full}

In this section we consider full QCD and present the finite volume
correlation
function of the scalar density operator $S=\bar{u} u+\bar{d}d+\bar{s}s$%
\bea \bra{0}| \pi^+_{-
\vec q}(t_1) \pi^-_{\vec q}(t_2) S(0)\ket{0} \, , \label{eq:uno}
\eea%
at NLO in the chiral expansion. We take $0<t_2\le t_1$. $\pi_{\vec
q}(t)$ is the Fourier transform of an interpolating operator for
the pion ($\pi(t,\vec
x\,)$)%
\beq \label{eq:interpol_op}
\pi_{\vec q}(t)\equiv  \int d^{\,3}x\mbox{ }{\mathrm{e}}^{-i \vec{q}
\cdot\vec{x}}
    \pi(t,\vec{x})\,.
\eeq
We only discuss correlation functions and amplitudes in the
centre-of-mass frame of the two final-state mesons.

\par We focus only on those terms which
concern the final state interactions (FSI). These are the terms
which generate finite-volume corrections to the matrix element
which decrease as powers of the volume and which give the shift in
the two-meson energy~\cite{luscher}. The remaining terms are only
subject to exponentially suppressed finite-volume corrections and
are not relevant in this discussion. The complete expressions for
the correlation functions at one-loop order in $\chi$PT can be
found in ref.~\cite{qi0}.
\par
At one-loop order in chiral perturbation theory, for an arbitrary
value of the two-pion energy, keeping only those terms which are
relevant for the energy shifts and finite-volume corrections
to the matrix elements, we find:%
\beq \bra 0|{\pi^+_{- \vec q}(t_1) \pi^-_{\vec q}(t_2) }
S(0)\ket{0} = \frac{e^{-Et_1}}{2E}\frac{e^{-Et_2}}{2E}
\l(-\frac{8}{f^2}\r)\l[1
\dots+I_b(t_{1},t_{2})\r]\, , \label{eq:scft} \eeq%
where $E=\sqrt{\vec q^{2} +m^{2}_{\pi}}$ and $f$ is the
pseudoscalar decay constant at lowest order in the chiral
expansion. We write $I_b(t_{1},t_{2})$ in the form%
\beq\label{eq:ib} I_b(t_{1},t_{2})= -\frac{E^2}{2
f^2}\frac{1}{L^3} \sum_{\vec k}\left(  d_{+}(w_\pi
)P(w_\pi)+c_{+}(w_K )P(w_K) + \frac{m_\pi^2}{6E^2 w_\eta
^2}P(w_\eta) \right) + \dots \, , \nn
\eeq where the volume is a cube of size $L^3$, and%
\bea
P(w)&=&\frac{1-e^{2(E-w)t_2}}{2(E-w)}\,, \quad w_{i}=\sqrt{\vec k^{2}
 +m^{2}_{i}}  \,, \nn \\
c_{+}(w)&=&\frac{2}{3}\frac{E^2 + Ew+w^2}{E^2w^2}\,, \quad
d_{+}(w)=2c_{+}(w)-\frac{m_\pi^2}{2E^2w^2}\,  . \nn \eea%
Evaluating the sum, we find
\begin{equation}
\bra 0|{\pi^+_{- \vec q}(t_1) \pi^-_{\vec q}(t_2)} S(0)\ket{0}
=\frac{e^{-Et_1}}{2E}\frac{e^{-Et_2}}{2E}\l(-\frac{8}{f^2}\r)\l[1+{\rm
Re}({\cal A}_{\infty})+{\cal T}(t_{2})\r]\,, \eeq %
where ${\cal A}_{\infty}$ is the infinite-volume one-loop
correction to the amplitude~\cite{qi0} and
\begin{equation}
{\cal T}(t_{2})=-\frac{E^2}{2f^2}\l\{ -\frac{\nu t_2}{E^2
L^3}(4-\frac{m_\pi^2}{2E^2} ) -\frac{2\nu t_2}{E^2
L^3}\theta(E-m_K) -\frac{\nu t_2}{E^2
L^3}\frac{m_\pi^2}{6E^2}\theta(E-m_\eta) \r\}+\dots \, , \nn
\label{eq:sopra} \eeq $\nu = \sum_{\vec k:w=E}$ and only the terms
which are proportional to $t_2$ are exhibited here.

The energy shift ($\Delta W$) can readily be extracted from the
terms proportional to $t_{2}$.  For $E<m_{K} < m_{\eta}$ only the
first term on the right-hand side of eq.~(\ref{eq:sopra})
contributes,
giving%
\beq \Delta W= -\l(2-\frac{m_\pi^2}{4E^2}\r)\frac{\nu}{f^2L^3} \,
. \label{eq:oes}  \eeq%
For energies such that $E > m_{\eta,K}=m_{\pi}$ on the other hand
we find%
\beq \Delta
W=-\l(3-\frac{m_\pi^2}{6E^2}\r)\frac{\nu}{f^2L^3} \, .
\label{eq:nes} \eeq%
The case $E < m_{K} < m_{\eta}$ was discussed in detail in
ref.~\cite{qi0} and we refer the reader to this reference for
details. We only remark that in this case only $P(w_\pi)$ in
eq.(\ref{eq:ib}) contains terms with vanishing denominators (and
hence terms proportional to $t_2$). Since we were considering
energies in the region $E < m_{K} < m_{\eta}$ in ref.~\cite{qi0},
the terms proportional to $\theta(E-m_K)$ and $\theta(E-m_\eta)$
were omitted in equation (23) of that paper, but here we also want
to study the case $m_{K} = m_{\pi} = m_{\eta} \leq E$, for which
$P(w_K)$ and $P(w_\eta)$ also have vanishing denominators, so that
the energy shift and the finite-volume corrections to the matrix
element are modified. This result is in agreement with the
L\"uscher quantization condition~\cite{luscher}, because the FSI
do depend on whether we are above or below the two-kaon (and
two-$\eta$) threshold. The consistency can be readily checked by
computing the one-loop expression of the relevant matrix element
in infinite volume in Minkowski space:
\begin{equation}
\bra{\pi^-(\vec q\,)\,\pi^+(-\vec q\,)}\,|\, S\, \ket{\,0\,}\equiv
-\frac{8}{f^2}\l[1+{\cal A}_{\infty}\r] =
-\frac{8}{f^2}\l[1+\frac{1}{(4\pi f)^2}\l( I_b+\dots \r)\r] \, ,
\label{eq:sft} \eeq where
\begin{eqnarray}
I_b&=&(m_\pi^2-2s)A(m_\pi)-s A(m_K)-\frac{m_\pi^2}{3}A(m_\eta)  +
\dots \,.\label{eq:spiega} \eea In the above expressions, $s =
(p_{\pi^-} +p_{\pi^+})^{2}$ is the square of the two-pion
centre-of-mass energy and
\begin{equation} A(m)\equiv
\sqrt{1-4\frac{m^2}{s}}\l(\ln{\frac{1+\sqrt{1-4\frac{m^2}{s}}}
{1-\sqrt{1-4\frac{m^2}{s}}}}-i\pi
\theta\l(1-4\frac{m^2}{s}\r)\r)\, . \label{eq:adm} \eeq %

When $E < m_{K} < m_{\eta}$, the s-wave phase shift, which is
obtained from the coefficient of the imaginary part of $I_{b}$, is
given by the first term of the r.h.s. of
eq.~(\ref{eq:spiega})~\footnote{ For $s \ge 4 m_\pi^2$, we have
${\rm Arg}\l[\bra{\pi^-(\vec q\,)\pi^+(-\vec q\,)}| S \ket{0}\r] =
\delta(s)$.}: \beq \delta(s)=\frac{2s-m_\pi^2}{16\pi
f^2}\sqrt{1-4\frac{m_\pi^2}{s}}\, . \eeq %

When $E > m_{K} = m_{\pi}= m_{\eta}$ on the other hand,  all the
terms in eq.~(\ref{eq:spiega}) contribute and we obtain
\begin{equation}
\delta(s)=\frac{9s- 2 m_\pi^2}{48\pi
f^2}\sqrt{1-4\frac{m_\pi^2}{s}}\, . \label{eq:ddegenerate}\eeq%
From the result for the phase shift in eq.~(\ref{eq:ddegenerate}),
we can extract the energy shift by using the L\"uscher
quantization formula~\cite{luscher}. The energy shift obtained in
this way agrees of course with the result in eq.~(\ref{eq:nes}),
obtained from a calculation of the correlation function in the
finite volume.

For the degenerate case $E > m_{K} = m_{\pi}= m_{\eta}$, the
energy shifts discussed here correspond to finite-volume
eigenstates which are two-meson $SU(3)$ singlet states ($S$ is an
$SU(3)$ singlet operator). A convenient procedure for the
evaluation of the matrix element is to compute the correlation
function of $S$ with the SU(3) singlet two-meson operator
$\bigl(\varphi\varphi\bigr)_{1}$,
\begin{equation}
S\textrm{\,-\,}\bigl(\varphi\varphi\bigr)_{1}\equiv\bra{0}|
\bigl(\varphi^{\dagger}_{-\vec q}(t_{1})\varphi_{\vec
q}(t_{2})\bigr)_{1} S(0)\ket{0}  \, , \label{eq:unos}
\end{equation}
where (for compactness of notation, in the following we suppress
the labels $t_{1,2}$ and $\vec q$) \beq
\bigl(\varphi\varphi\bigr)_{1} = \frac{1}{\sqrt{2}}\, KK  +
\frac{1}{2\sqrt{2}} \,  \eta \eta +
\frac{1}{2}\sqrt{\frac{3}{2}}\,  \pi\pi \, ,  \eeq and \bea
 KK  &=& \frac{1}{\sqrt{2}} \left(  K^{0} \bar K^{0}
+  K^{+}K^{-}  \right) \, , \quad \quad   \pi\pi =
\frac{1}{\sqrt{3}} \left(  \pi^{0}  \pi^{0} + \sqrt{2}
\pi^{+}\pi^{-}\right) \, . \label{eq:osinglet} \eea
$(\varphi\varphi\bigr)_{1}$  creates or annihilates the SU(3)
singlet state $\vert 1 \rangle$, which is a combination of
two-pion, two-kaon and two-$\eta$ states with the same flavour
structure as the corresponding operator
\begin{eqnarray} \vert 1
\rangle &=& \frac{1}{\sqrt{2}} \vert KK \rangle +
\frac{1}{2\sqrt{2}} \vert \eta \eta\rangle
+\frac{1}{2}\sqrt{\frac{3}{2}} \vert \pi\pi \rangle \, , \nn \\
\vert KK \rangle &=& \frac{1}{\sqrt{2}} \left( \vert K^{0} \bar
K^{0} \rangle+ \vert K^{+}K^{-} \rangle \right) \, , \quad \quad
\vert \pi\pi \rangle = \frac{1}{\sqrt{3}} \left( \vert \pi^{0}
\pi^{0} \rangle+ \sqrt{2} \vert \pi^{+}\pi^{-} \rangle \right) \,
. \label{eq:singlet} \eea All the meson states are $s$-wave, $I=0$
states. Similar expressions can be written for the octet and
27-plet operators considered below. Computing in addition the
$(\varphi\varphi\bigr)_{1}$\,-\,$(\varphi\varphi\bigr)_{1}$
correlation function \bea \langle 0\vert
\bigl(\varphi^{\dagger}_{-\vec q}(t_{1})\varphi_{\vec
q}(t_{2})\bigr)_{1} \bigl(\varphi^{\dagger}_{\vec
q}(-t_{2})\varphi_{-\vec q}(-t_{1})\bigr)_{1} \vert 0\rangle\,,
\label{eq:dues} \eea one obtains the finite volume matrix element
$\vert \langle 1  \vert S \vert  0 \rangle\vert_{FV}$ by dividing
the  correlation function (\ref{eq:unos}) by the square root of
the four-point function (\ref{eq:dues}) at large time distances,
following the procedure explained in ref.~\cite{qi0} \bea  \vert
\langle 1 \vert S \vert  0 \rangle\vert_{FV} =\frac{\bra{0}|
\bigl(\varphi^{\dagger}_{-\vec q}(t_{1})\varphi_{\vec
q}(t_{2})\bigr)_{1} S(0)\ket{0} }{\sqrt{\langle
0\vert\bigl(\varphi^{\dagger}_{-\vec q}(t_{1})\varphi_{\vec
q}(t_{2})\bigr)_{1} \bigl(\varphi^{\dagger}_{\vec
q}(-t_{2})\varphi_{-\vec q}(-t_{1})\bigr)_{1}\vert 0 \rangle }}\,
.\label{eq:sphiphi} \eea The $\pi\pi$ matrix element is then
simply obtained by using the appropriate Clebsh-Gordan
coefficient, $\langle \pi\pi \vert S \vert  0 \rangle = 2
\sqrt{2/3} \langle 1 \vert S \vert 0 \rangle$.

For the sake of presentation, it was convenient to discuss the
extraction of the matrix element using the ratio on the right-hand
side of eq.~(\ref{eq:sphiphi}), with the correlation function
$S\textrm{\,-\,}\bigl(\varphi\varphi\bigr)_{1}$ in the numerator.
Note however, that since $S$ is an $SU(3)$-singlet, we could
equally well have used the correlation function
$S\textrm{\,-\,}\bigl(\pi\pi\bigr)$.

In the case of $(8,1)$ or $(8,8)$ operators the simple procedure
outlined above does not work because these operators have
non-vanishing matrix elements between the kaon state, $\vert K^{0}
\rangle$, and three different S-wave two-meson states with $I=0$
and $I_{z}=0$. The three states are the state $\vert 1 \rangle$
defined above and the octet and 27-plet states (created by the
corresponding $(\varphi\varphi\bigr)_{8}$ and
$(\varphi\varphi\bigr)_{27}$ operators) \bea \vert 8 \rangle &=&
\frac{1}{\sqrt{5}} \vert KK \rangle + \frac{1}{\sqrt{5}} \vert
\eta \eta\rangle
-\sqrt{\frac{3}{5}} \vert \pi\pi \rangle  \, , \nn \\
\vert 27 \rangle &=& -\sqrt{\frac{3}{10}} \vert KK \rangle +
\sqrt{\frac{27}{40}} \vert \eta \eta\rangle +\frac{1}{\sqrt{40}}
\vert \pi\pi \rangle \, . \eea In a finite volume the three
two-meson states ($\vert 1 \rangle$, $\vert 8 \rangle$ and $\vert
27 \rangle$) acquire different energies,  and thus they appear in
the correlation function with different exponentials in time.
Thus, in order to obtain the matrix element $\langle \pi\pi \vert
Q_{6} \vert  K^{0} \rangle$,  for example, one has to disentangle
the different contributions, and  extract all the matrix elements
$\langle 1,8,27 \vert  Q_{6} \vert  K^{0} \rangle$ and $\langle
1,8,27  \vert  1,8,27  \rangle$ by studying the
$K^{0}$-$Q_{6}$-$(\varphi\varphi\bigr)_{i}$ and the
$(\varphi\varphi\bigr)_{i}$-$(\varphi\varphi\bigr)_{i}$
correlators.  Then a suitable combination of $\langle 1,8,27 \vert
Q_{6} \vert  K^{0} \rangle$ will give the required $\langle \pi\pi
\vert  Q_{6} \vert  K^{0} \rangle$ amplitude. A further
complication is that all the matrix elements should be computed at
the same two-meson energy, which must be kept fixed in the
infinite volume limit~\cite{LL,noi}. Although possible in
principle, the procedure appears to be very complicated to
implement in practice.

\section{Two-pion matrix elements in partially quenched QCD}
\label{sec:partially}

In this section we discuss the evaluation of the same matrix
elements, but now in partially quenched QCD. We consider the two
cases: QCD with three sea-quark flavours for generic values of
$m_{S}$, $m_{V}$ and $m_{s}$ (PQ3), and QCD with two light
sea-quark flavours also with generic combinations of the quark
masses (PQ2). These calculations correspond to theories with
$SU(6|3)$ and $SU(5|3)$ graded Lie groups respectively. We have
also studied the $SU(4|2)$ case, reaching the same physical
conclusions, and for this reason this case will not be discussed
explicitly.

PQ2 is partially quenched two-flavour QCD, with a mass $m_{S}$ for
the sea quarks, a mass $m_{V}$ for the valence light quarks and a
mass $m_{s}$ for the strange quark, which is always quenched. The
rather lengthy formulae for $m_{S} \neq m_{V}$ are presented in
the appendix for both the scalar operator and (8,1) operators. We
explicitly confirm the expectations~\cite{Laiho:2003uy} that when
the sea-quark mass is different from the valence-quark mass the
problems are the same as in the quenched case~\cite{qi0}. The
double poles induce terms which seem to grow linearly and
cubically with the volume as well as ones which are quadratic or
cubic in the time distance, see eq.~(\ref{eq:mammasantissima}).
The extraction of the physical amplitude is therefore not
possible, at least with our current level of understanding. In
this respect we find that it is euphemistic to state that ``the
final state corrections can be significant''~\cite{Laiho:2003uy};
we simply do not know how to extract the infinite volume matrix
element from the finite volume correlation function and the
problem cannot be overcome by going to larger lattices.  This
pathology is not peculiar to PQ2 but is also present with the same
symptoms in PQ3.

For $m_{S} = m_{V}$, for a generic energy $E$ and a generic
strange quark mass $m_{s}$ we obtain in PQ2:
\begin{equation} \bra{0}| \pi^+_{- \vec
q}(t_1)\pi^-_{\vec q}(t_2)S^{pq}(0) \ket{0} =
\frac{e^{-Et_1}}{2E}\frac{e^{-Et_2}}{2E} \l(-\frac{8}{f^2}\r)\l[1
\dots +I^{pq}_b(t_1,t_2)\r]\, , \eeq where we choose
$S^{pq}=\bar{u}u+\bar{d}d+\bar{s}s$ and the sum is only over the
valence quarks,
\begin{equation}
I^{pq}_b(t_1,t_2)=-\frac{E^2}{2 f^2} \frac{1}{L^3}
\sum_{\vec k} \l( d_{+}(w_{\pi}) P(w_{\pi})  +\frac{1}{2}
c_{+}(w_{K}) P(w_{K})+ \frac{m_\pi^2}{4E^2 w_{\bar s s}^2}
P(w_{\bar s s }) \r) + \dots \nn \eeq and $m^{2}_{\bar s s }= 2
m_{K}^{2}-m_{\pi}^{2}$. When $m_{K} > m_{\pi}$ ($m_{s} > m_{V} =
m_{S}$) and taking the two-pion energy such that $W_{\pi}<
W^{min}_{K}$ (at this order in the chiral expansion this means $2
E \leq 2 m_{K}$), only the term proportional to $P(w_{\pi})$
generates the energy shift and the finite volume power
corrections, which are exactly the same as in the unquenched
two-flavour theory. In this respect we agree with the conclusions
of LS for $m_{K} = 2 m_{\pi}$. More generally, one has finite
volume corrections under control and the validity of the LL
formula, as in two-flavour full QCD, whenever $W_{\pi} < 2 m_{K}$.
Indeed, $W_{\pi} <  2 m_{K}$ is precisely the elasticity condition
under which the LL formula applies. If we violate this condition
however, then major difficulties arise. Consider for example the
case $m_{K} = m_{\pi} = m_{\bar s s }$: each of the three terms
contributes to the energy-shift, and the final coefficient does
not correspond to the correct result either in SU(2) or in SU(3).
The reason for this is that the ``quenched'' states containing a
strange quark now also contribute to the singular terms in the
sum. The matrix elements now correspond to final states with a
ghost component and therefore we are unable to extract the
required two-pion amplitudes. In full QCD in sec.~\ref{sec:full},
we were able to use the SU(3) symmetry in the $m_u=m_d=m_s$ case
to determine the $\pi\pi$-component in an SU(3)-singlet state.
Here we are unable to carry out the analogous procedure. We note
that one could avoid this difficulty by using a scalar operator
which projects only onto two-pion states, such as
$\bar{u}u+\bar{d}d$. In this case the problem of the mixing with
intermediate states containing strange quarks does not arise. This
feature, however, does not apply to the $(8,1)$ or $(8,8)$
operators of the weak effective Hamiltonian.

The problems above, which are related to the absence of unitarity
in partially quenched theories, are also present in PQ3 when
$W_\pi$ is above the threshold for any multi-hadron state 
containing ``partially-quenched" quarks. Therefore in order to
extract the $\Delta I=1/2$ $K\to\pi\pi$ matrix elements with
$m_u=m_d=m_s$ in PQ3, one is forced to work with the full QCD
limit of the theory. 

\section{Conclusions}\label{sec:concs}
In this paper we study the extraction of matrix elements with
two-pion final states from simulations in finite volumes in full
and partially quenched QCD. In full QCD, not surprisingly, we
confirm that the matrix elements can be determined in principle,
up to exponentially small finite-volume corrections. We point out
however, that even in this case, for three degenerate quark
flavours there are considerable practical difficulties because the
finite-volume energy eigenstates correspond to linear combinations
of $\pi\pi$, $KK$ and $\eta\eta$ states. The determination of a
matrix element with a two-pion final state with a fixed energy
requires therefore the extraction of several finite-volume matrix
elements at this energy. Each of these matrix elements will, in
general, need to be computed from a simulation on a different
volume.

For PQ2 we find that it is not possible to determine $K\to\pi\pi$
matrix elements in general, at least with our present level of
understanding. For two-pion energies below the two-kaon threshold
however, the theory is a consistent unquenched two-flavour one,
the LL formula holds and it is possible to extract infinite volume
physical amplitudes. Of course the value of the couplings of the
effective weak chiral Lagrangian and the coefficients of the
logarithms which appear at one loop are not those of full QCD. One
may argue, however, that the contribution to these couplings due
to the sea-strange quark loops are suppressed by the massive kaon
propagators when $W_{\pi} << 2 m_{K}$, since there are no
singularities associated with the strange meson loops. If this is
true, PQ2 can provide very useful physical information, although
we are currently unable to estimate the systematic uncertainty.
Given the apparent importance of unitarity in enabling the
determination of physically meaningful results, we believe that it
may be theoretically better to work with the unquenched two
flavour theory rather than to weight the gauge field
configurations with fractional powers of the fermion determinant
in order to mimic the three flavour theory. A study of this issue
in chiral perturbation theory may be very useful to clarify the
situation.

The disappointing news is the requirement that the sea and valence
quark masses should be equal. It had been considered very useful
to work with fixed sea-quark masses and to vary (and in particular
to reduce) the valence ones, in order to calibrate the lattice
spacing and to extrapolate the results to the physical point. In
this respect, our results cast serious doubts on the applicability
of PQ3 to the determination of matrix elements with two-meson (or
in general, multi-hadron) external states. PQ3 may however, prove
useful for the determination of some other quantities.

For PQ2 with the two-pion energy at or above the two-kaon
threshold (more precisely for $m_{K} \ge  m_{\pi}$ 
($m_{s} \ge m_{V} = m_{S}$), with
$W_{\pi}\ge W^{min}_{K}$) with our present level of understanding
it is not possible to extract the two-pion matrix elements. The
difficulties are those already discussed for quenched
QCD~\cite{qi0}. Unfortunately this implies that the proposal
suggested by LS of combining $K \to 0$, $K \to \pi$ and $K \to \pi
\pi$ amplitudes, with the two pions at rest, to determine the
complete set of couplings of the weak chiral Lagrangian fails in
PQ2. An ingredient in the LS proposal is the extraction of the
two-pion matrix element with $m_\pi=m_K$, which violates the
consistency condition given above. Of course the LS proposal may
work in full QCD, but requires care in the extraction of the $K
\to \pi\pi$ matrix element when $m_{K} = m_{\pi}$. In this case
the full two meson $\to $ two meson correlation matrix must be
studied, as explained in sec.~\ref{sec:full}.

\section*{Acknowledgements}
We thank Jonathan Flynn and Steve Sharpe for helpful discussions
and correspondence.

This work was supported by European Union grant
HTRN-CT-2000-00145. CJDL and CTS acknowledge support from PPARC
through grants PPA/G/S/1998/00529 and PPA/G/O/2000/00464. EP was
supported in part by the Italian MURST under the program
\textit{Fenomenologia delle Interazioni Fondamentali}.

\appendix

\section{Finite-volume correlation functions in partially quenched QCD}
\label{su53ndfv}

In this appendix we present the results for the correlation
functions used in the extraction of two-pion matrix elements in
partially quenched QCD, PQ2.  The Euclidean partially quenched
chiral Lagrangian in the supersymmetric formulation
is~\cite{Bernard:1993sv}
\begin{eqnarray}
 {\mathcal{L}}^{pq}_{\chi} &=& \frac{f^{2}}{8}{\mathrm{str}}
 \left [ \left ( \partial_{\mu} \Sigma^{pq\dagger} \right )
  \left ( \partial_{\mu} \Sigma^{pq} \right )\right ]
 - \frac{f^{2}}{8} {\mathrm{str}}\left [
  \Sigma^{pq\dagger}\chi +
  \chi^{\dagger}\Sigma^{pq}\right ]
 - m_{0}^{2}\Phi^{2}_{0} + \alpha \left ( \partial_{\mu}\Phi_{0}\right )
  \left ( \partial_{\mu}\Phi_{0}\right ) ,
\end{eqnarray}
where $\Sigma^{pq}$ is the graded extension of the standard
non-linear Goldstone field $\Sigma$ in the full chiral
perturbation theory, $\Phi_{0}$ is the super-$\eta^{\prime}$ field
and $\chi=2B_0{\cal M}$, where ${\cal M}$ is the mass matrix and
$B_0=-\langle0|\bar{u}u+\bar{d}d|0\rangle/f^2$.

The results reported here are obtained with the
super-$\eta^{\prime}$ integrated out (the limit
$m_{0}\rightarrow\infty$), and are presented in terms of the
following quantities:
\begin{eqnarray}
c_{\pm}(w)&=&\frac{2}{3}\frac{E^2 \pm Ew+w^2}{E^2w^2}\,;\quad
c_{0}(w)=\frac{2}{3}\frac{1}{E^2}\,;
    \quad d_{\pm,0}(w)=2c_{\pm,0}(w)-\frac{m_\pi^2}{2E^2w^2}\, ;\nn \\
w_{S}&=&\sqrt{\vec k^{2} +m^{2}_{VS}}\,, \quad
m_{VS}^2=B_0(m_{u}+m_{S})\qquad\textrm{and}\label{eq:defs}\\ %
w_{i}&=&\sqrt{\vec k^{2} +m^{2}_{i}}\quad (\textrm{for\ }i\neq S),
\quad \delta_s=(m_\pi^2-m_{VS}^2)\, , \quad
B_{ij}=\frac{m_i^2-m_{VS}^2}{m_i^2-m_j^2}\,. \nn
\end{eqnarray}
The finite volume scalar correlation function can be written in
the following form ($E=\sqrt{\vec{q}^{\ 2}+m_\pi^2}$): \beq
\bra{0}| \pi^+_{- \vec q}(t_1)\pi^-_{\vec q}(t_2)S^{pq}(0) \ket{0}
= \frac{e^{-Et_1}}{2E}\frac{e^{-Et_2}}{2E}
\l(-\frac{8}{f^2}\r)\l[1+
 I^{pq}_z + I^{pq}_a+I^{pq}_b(t_1,t_2)\r]\, ,
\end{equation}
where  $I^{pq}_z$ and $I^{pq}_a$ are the contributions from the
wave-function renormalization and tadpole diagram respectively.
They give the infinite volume result up to exponentially
suppressed corrections with the volume and are therefore not
reported explicitly here.
\begin{equation}
I^{pq}_b(t_1,t_2)=-\frac{E^2}{2
f^2}\l[A_{SS}+A_{KK}+A_{\pi\pi}+A_{0\pi}+A_{00}\r] \, , \nn \eeq
where
\begin{eqnarray}
A_{SS}&=&\frac{1}{L^3} \sum_{\vec
k}\l\{\frac{c_{+}(w_S)}{2(E-w_S)}-\frac{c_{-}(w_S)}{2(E+w_S)}
        -e^{2(E-w_S)t_2}\l(\frac{c_{+}(w_S)}{2(E-w_S)}+\frac{c_{0}(w_S)}
{2w_S}\r)\r.  \nn \\
    &&\quad\l.+e^{2Et_2-2w_St_1}\l(\frac{c_{0}(w_S)}{2w_S}-
\frac{c_{-}(w_S)}{2(w_S+E)}\r)\r\}\, ,  \nn \\
A_{KK}&=&\frac{1}{L^3} \sum_{\vec k}\frac12
\l\{\frac{c_{+}(w_K)}{2(E-w_K)}-\frac{c_{-}(w_K)}{2(E+w_K)}
        -e^{2(E-w_K)t_2}\l(\frac{c_{+}(w_K)}{2(E-w_K)}+
\frac{c_{0}(w_K)}{2w_K}\r)\r.  \nn \\
    &&\quad \l.+e^{2Et_2-2w_Kt_1}\l(\frac{c_{0}(w_K)}{2w_K}-
\frac{c_{-}(w_K)}{2(w_K+E)}\r)\r\}\, , \nn \\
A_{\pi\pi}&=&\frac{1}{L^3} \sum_{\vec
k}\l\{\frac{d_{+}(w_{\pi})-c_{+}(w_{\pi})}
{2(E-w_{\pi})}-\frac{d_{-}(w_{\pi})-c_{-}(w_{\pi})}{2(E+w_{\pi})}\r.  \nn \\
    &&\quad \l. -e^{2(E-w_{\pi})t_2}\l(\frac{d_{+}(w_{\pi})
-c_{+}(w_{\pi})}{2(E-w_{\pi})}
    +\frac{d_{0}(w_{\pi})-c_{0}(w_{\pi})}{2w_{\pi}}\r) \r.  \nn \\
    &&\quad \l. +e^{2Et_2-2w_{\pi}t_1}\l(\frac{d_{0}(w_{\pi})
-c_{0}(w_{\pi})}{2w_{\pi}}
    -\frac{d_{-}(w_{\pi})-c_{-}(w_{\pi})}{2(w_{\pi}+E)}\r)\r\} \nn \\
    &&+\frac{1}{L^3} \sum_{\vec k}\frac{m_\pi^2 
(2\delta_s^2+w_\pi^4 B_{\pi K}^2)}
    {4E^2w_{\pi}^6}\l\{\frac{1}{2(E-w_{\pi})}-\frac{1}{2(E+w_{\pi})}
-e^{2(E-w_{\pi})t_2}\l(\frac{1}{2(E-w_{\pi})}+\frac{1}{2w_{\pi}}\r)\r. 
 \label{eq:mammasantissima} \\
    &&\quad \l.+e^{2Et_2-2w_{\pi}t_1}\l(\frac{1}{2w_{\pi}}-\frac{1}
{2(w_{\pi}+E)}\r)\r\} \nn  \\
    &&+\frac{1}{L^3} \sum_{\vec k}-\frac{m_\pi^2 \delta_s^2}{E^2w_{\pi}^5}
    \l\{\frac{1}{[2(E-w_{\pi})]^2}+\frac{1}{[2(E+w_{\pi})]^2}
        -e^{2(E-w_{\pi})t_2}\l(\frac{1-[2(E-w_{\pi})] t_2}{[2(E-w_{\pi})]^2}
        -\frac{1+[2w_{\pi}]t_2}{[2w_{\pi}]^2}\r)\r.  \nn \\
    &&\quad \l.-e^{2Et_2-2w_{\pi}t_1}\l(\frac{1+[2w_{\pi}]t_1}{[2w_{\pi}]^2}
    -\frac{1+[2(w_{\pi}+E)]t_1}{[2(w_{\pi}+E)]^2}\r)\r\} \nn  \\
    &&+\frac{1}{L^3} \sum_{\vec k}\frac{m_\pi^2 \delta_s^2}{E^2w_\pi^4}
    \l\{\frac{1}{[2(E-w_{\pi})]^3}-\frac{1}{[2(E+w_{\pi})]^3} \r. \nn \\
    &&\quad \l.-e^{2(E-w_{\pi})t_2}\l(\frac{1-[2(E-w_{\pi})] 
t_2+[2(E-w_{\pi})]^2 t_2^2/2}{[2(E-w_{\pi})]^3}
        +\frac{1+[2w_{\pi}]t_2+[2w_{\pi}]^2 t_2^2/2}{[2w_{\pi}]^3}\r)\r. \nn \\
    &&\quad \l.+e^{2Et_2-2w_{\pi}t_1}\l(\frac{1+[2w_{\pi}]t_1+
[2w_{\pi}]^2t_1^2/2}{[2w_{\pi}]^3}
        -\frac{1+[2(w_{\pi}+E)]t_1+[2(w_{\pi}+E)]^2t_1^2/2}
{[2(w_{\pi}+E)]^3}\r)\r\}\, ,\nn \\
A_{0\pi}&=&\frac{1}{L^3} \sum_{\vec k}\frac{m_\pi^2 B_{\pi K}B_{K
\pi}}{2E^2 w_{\pi} w_{\bar s s}}\l\{\frac{1}{2E-w_{\pi}-w_{\bar s
s}}-\frac{1}{2E+w_{\pi}+w_{\bar s s}}
        -e^{(2E-w_{\pi}-w_{\bar s s})t_2}\l(\frac{1}{2E-w_{\pi}-w_{\bar s s}}+
        \frac{1}{w_{\pi}+w_{\bar s s}}\r)\r.  \nn \\
    &&\quad \l.+e^{2Et_2-(w_{\pi}+w_{\bar s s})t_1}
\l(\frac{1}{w_{\pi}+w_{\bar s s}}-
    \frac{1}{w_{\bar s s}+w_{\pi}+2E}\r)\r\} \nn \, ,  \\
A_{00}&=&\frac{1}{L^3} \sum_{\vec k}\frac{m_\pi^2 B_{K
\pi}^2}{4E^2 w_{\bar s s}^2} \l\{\frac{1}{2(E-w_{\bar s
s})}-\frac{1}{2(E+w_{\bar s s})}
        -e^{2(E-w_{\bar s s})t_2}\l(\frac{1}{2(E-w_{\bar s s})}
+\frac{1}{2w_{\bar s s}}\r)\r.  \nn \\
    &&\quad \l.+e^{2Et_2-2w_{\bar s s}t_1}\l(\frac{1}{2w_{\bar s s}}
-\frac{1}{2(w_{\bar s s}+E)}\r)\r\} \, . \nn
\eea Note that the presence of flavour-singlet double poles gives
rise to terms proportional to powers of $t_1$ and $t_2$ and that
these terms disappear when $m_S=m_{V}$ (when $\delta_s$ and
$B_{\pi K}$ vanish).

The same structure is present in the $K\to\pi\pi$ correlation
functions of the (8,1) operator. We denote this operator by
$Q_{(8,1)}^{pq}$ ($Q_6$ is an example of a QCD operator which
transforms as the (8,1) representation), and in the chiral
effective theory it is chosen to be
\bea
 Q_{(8,1)}^{pq} &=& {\mathrm{str}} \left [ \lambda^{pq}_{6}
  \partial_{\mu} \Sigma^{pq}\partial_\mu\Sigma^{pq\dagger}\right ] ,
\qquad\textrm{where}\nonumber\\
 \lambda^{pq}_{6} &=& \left ( \begin{array}{cc}
 \lambda_{6} & 0\\ 0 & 0\end{array} \right ) .
\eea The correlation function is given by
\beq \bra{0}| \pi^+_{- \vec q}(t_1)\pi^-_{\vec
q}(t_2)Q_{(8,1)}^{pq}(0) {\bar K}^0_{\vec 0} (t_K)\ket{0} =
\frac{e^{-Et_1}}{2E}\frac{e^{-Et_2}}{2E}\frac{e^{m_K t_K}}{2m_K}
\l(-\frac{8 i}{f^3}\r)\l[
\tau(E,m_K,m_\pi)-\frac{E^2}{2f^2}\l(\sum_{I=\pi\pi, KK, \dots}
A_{I} + ... \r)\r]\, ,
\eeq where the ellipses represent non-singular contributions, the
sum runs over all mesons and
$$\tau(E,m_1,m_2)\equiv \frac{2E^2+Em_1-m_2^2}{2}\,.$$
To this order the relevant contributions to $A_{I}$ can be written
as \beq A_{I} = \frac{1}{L^3} \sum_{\vec k} \sum_{r=1}^{3}
\frac{{\cal C}^{(r)}_{I}}{(2E-w_i-w_j)^r}
\l(1-e^{(2E-w_i-w_j)t_2}\sum_{n=0}^{r-1}
(2E-w_i-w_j)^n\frac{t_2^n}{n!} \r) + ... \, , \nn \eeq where again
only the terms containing poles have been kept. The only non-zero
coefficients are the following: \bea
{\cal C}^{(1)}_{KK}&=&-\tau(w_K,m_K,m_K) c_+(w_K) \, , \nn \\
{\cal C}^{(1)}_{SS}&=&\tau(w_S,m_K,m_{VS}) c_+(w_S) \, , \nn \\
{\cal C}^{(1)}_{\pi\pi}&=&\tau(w_\pi,m_K,m_\pi) (d_+(w_\pi)-c_+(w_\pi))\nn \\
&&+\frac{m_\pi^2}{4E^2w_\pi^6}\l( \delta_s ^2
(w_\pi^2-m_\pi^2)-m_Kw_\pi^3\delta_sB_{\pi K}-2w_\pi^4 \,
\tau(w_\pi,m_K,m_\pi) B_{\pi K}^2\r) \, , \nn \\
{\cal C}^{(2)}_{\pi\pi}&=&-\delta_s^2  \frac{m_\pi^2}{4E^2w_\pi^5}
\l(2w_\pi^2+w_\pi m_K-2m_\pi^2\r) \, , \nn \\
{\cal C}^{(3)}_{\pi\pi}&=& \delta_s^2 \tau(w_\pi,m_K,m_\pi)
 \frac{m_\pi^2}{E^2w_\pi^4} \, , \nn \\
{\cal C}^{(1)}_{0\pi}&=& -B_{\pi
K}B_{K\pi}\frac{m_\pi^2}{E^2w_{\bar s s}w_\pi}\,\tau\l
(\frac{w_{\bar s s}+w_\pi}{2},m_K,m_K\r)-
\frac{m_\pi^2m_K \delta_s B_{K\pi}}{4E^2w_\pi^3}\, , \nn \\
{\cal C}^{(2)}_{0\pi}&=& \frac{m_\pi^2m_K\delta_s B_{K\pi}
(w_{\bar s s}-w_\pi)}{4E^2w_{\bar s s}w_\pi^2} \, , \nn \\
{\cal C}^{(1)}_{00}&=&-B_{K \pi}^2\,\tau\l(w_{\bar s
s},m_K,\sqrt{2m_K^2-m_\pi^2}\r) \frac{m_\pi^2}{2E^2w_{\bar s s}^2}
\, . \nn \eea where the quantities on the right-hand side are
defined in eq.~(\ref{eq:defs}).


\end{document}